\documentstyle[11pt]{article}
\begin{document}
\newcommand{\half}{\mbox{\small{$\frac{1}{2}$}}}
\renewcommand{\refname}{References.}
\title{Gauged Nambu--Jona-Lasinio model at $O(1/N)$ with and without a
Chern-Simons term.}
\author{J.A. Gracey, \\ Department of Applied Mathematics and Theoretical
Physics, \\ University of Liverpool, \\ P.O. Box 147, \\ Liverpool, \\
L69 3BX, \\ United Kingdom.}
\date{}
\maketitle
\vspace{5cm}
\noindent
{\bf Abstract.} We solve the gauged Nambu--Jona-Lasinio model at leading
order in the large $N$ expansion by computing the anomalous dimensions of all
the fields of the model and other gauge independent critical exponents by
examining the scaling behaviour of the Schwinger Dyson equation. We then
restrict to the three dimensional model and include a Chern Simons term to
discover the $\theta$-dependence of the same exponents where $\theta$ is the
Chern Simons coupling.

\vspace{-16cm}
\hspace{10cm}
{\bf LTH-311}
\newpage
Recently there has been a resurgence of interest in examining the dynamics of
models with four fermi interactions due to the possible composite nature of the
Higgs boson of the standard model, \cite{1,2}. This is based partly on the
observation that, similar to what occurs in the Gross Neveu model \cite{3},
fermion pairs can bind together to form a bosonic particle whose mass is around
twice that of one of the fermions. In \cite{1,2} it was discussed
how this bound state can effect the spontaneous symmetry breaking of the Higgs
boson in the more conventional understanding of the generation of mass for the
particles we observe in the real world. To improve our knowledge of the basic
four fermi model which underlies this alternative approach it is important to
have a comprehensive picture of the quantum properties of such models. One
technique widely used to probe these theories is the large $N$ expansion. As a
first step in this direction the $O(N)$ Gross Neveu model has recently been
solved at $O(1/N^2)$ in a $1/N$ approximation where $N$ is the number of
fundamental fermions, \cite{4,5}. The method used was based on the earlier work
of \cite{6,7} and involved computing the anomalous dimensions of the basic
fields and bound states of the theory in {\em arbitrary} dimensions by solving
the theory near its $d$-dimensional critical point. There the fields obey a
conformal symmetry and their critical exponents, which are related to the
critical renormalization group functions, can be deduced by solving the
Schwinger Dyson equations self consistently. More recently the chiral Gross
Neveu or Nambu--Jona-Lasinio model, \cite{3}, has also been solved at
$O(1/N^2)$ in the same approach, \cite{8}. However, to fully understand the
dynamics of these simplified models in relation to the standard model it is
important to consider the more realistic scenario where four fermi interactions
are coupled to gauge fields. In this letter we report the results of such an
analysis for the Nambu--Jona-Lasinio model coupled to QED where we have a
continuous global $U(1)$ $\times$ $U(1)$ chiral symmetry. We will deduce the
anomalous dimensions of the fields in arbitrary dimensions in a large $N$
expansion. This will be important for an $\epsilon$-expansion approach to
understanding which universality class the models lie in and for comparison
with the perturbation theory of other models.

As we will deal with a model with a $U(1)$ gauge field it is also important to
record new results for another area where four fermi models are also of
interest. It has recently been argued that a model which underlies high
$T_c$ superconductivity involves a four fermi interaction coupled to QED in
three dimensions, \cite{9}. Whilst our results will produce three dimensional
gauge independent exponents, which are therefore physically relevant, we will
also couple our model explicitly to a Chern Simons term specifically in three
dimensions, [10-12]. The motivation for including such a topological term is
its relation to anyons which are believed to be a mechanism for high $T_c$
superconductivity, \cite{13,14}. Therefore, it is important to produce gauge
independent exponents which depend on the Chern Simons coupling $\theta$, which
does not get renormalized in the quantum theory and therefore appears as a pure
$N$-independent parameter in the exponents, \cite{A}. As the gauge independent
exponents are physically interesting and measurable, this will allow one to
determine whether $\theta$ is non-zero or not and therefore if such topological
terms exist in nature.

The lagrangian of the theory we consider is
\begin{eqnarray}
L &=& i \bar{\psi}^i{\partial \!\!\! /} \psi^i + \sigma \bar{\psi}^i \psi^i
+ i\pi \bar{\psi}^i\gamma^5\psi^i + A_\mu \bar{\psi}^i\gamma^\mu\psi^i
\nonumber \\
&-& \frac{1}{2g^2}(\sigma^2 + \pi^2) - \frac{1}{4e^2} F^2_{\mu\nu}
\end{eqnarray}
where $\sigma$ and $\pi$ are auxiliary bosonic fields whose elimination yields
the usual four fermi interaction and which are fermion bound states in the
quantum theory, \cite{3}. We have scaled the coupling constant $e$ and $g$ in
such a way that the couplings of the $3$-vertices are unity and $A_\mu$ is our
$U(1)$ gauge field. The fields $\psi^i$ lie in an $N$-tuplet, $1$ $\leq$ $i$
$\leq$ $N$, and $1/N$ will be our expansion parameter. We follow the
conventions of \cite{15} in defining the properties of $\gamma^5$ as $\{
\gamma^\mu, \gamma^5\}$ $=$ $0$ and $\mbox{tr}(\gamma^5\gamma^{\mu_1}...
\gamma^{\mu_n})$ $=$ $0$, where we take $\mbox{tr}1$ $=$ $4$.

To solve (1) via the self consistency approach of [4-7] the anomalous
dimensions of the fields are deduced by substituting the asymptotic scaling
forms of the fields at criticality into their skeleton Dyson equations with
dressed propagators. More concretely for (1), we take in coordinate space,
as $x$ $\rightarrow$ $0$, \cite{4,16},
\begin{eqnarray}
\psi(x) & \sim & \frac{A {x \!\!\! /}}{(x^2)^\alpha} ~~,~~
\sigma(x) ~ \sim ~ \frac{B}{(x^2)^\beta} ~~,~~
\pi(x) ~ \sim ~ \frac{C}{(x^2)^\gamma} \nonumber \\
A_{\mu\nu}(x) & \sim & \frac{D}{(x^2)^\rho} \left[ \eta_{\mu\nu}
+ \frac{2\rho(1-b)}{(2\mu-2\rho-1+b)} \frac{x_\mu x_\nu}{x^2} \right]
\end{eqnarray}
where $b$ is the covariant gauge parameter. Retaining $b$ in our calculation
will prove useful in demonstrating which of our exponents are gauge independent
and therefore physically relevant. The quantities $A$, $B$, $C$ and $D$ are the
respective amplitudes of the fields and $\alpha$, $\beta$, $\gamma$ and $\rho$
are the dimensions of the fields and we define their anomalous dimensions by
\begin{equation}
\alpha ~=~ \mu + \half \eta ~,~ \beta ~=~ 1 - \eta - \chi_\sigma ~,~
\gamma ~=~ 1 - \eta - \chi_\pi ~,~ \rho ~=~ 1 - \eta - \chi_A
\end{equation}
where we define the spacetime dimension to be $d$ $=$ $2\mu$, $\eta$ is the
fermion anomalous dimension and $\chi$ denotes the vertex anomalous dimension
of the respective $3$-vertices of (1) and they contain information on the
structure of the quantum theory. To compute each of these, which depend on $N$,
$\mu$ and are $O(1/N)$, one represents the skeleton Dyson equations with
dressed propagators, which are valid in the critical region $x$ $\rightarrow$
$0$, by the scaling forms (2). At leading order these are illustrated in fig. 1
and therefore they correspond to
\begin{eqnarray}
0 &=& r(\alpha-1) + z + y + \frac{2u[\rho(1-b)-(\mu-1)(2\mu-2\rho-1+b)]}
{(2\mu-2\rho-1+b)} \\
0 &=& p(\beta) + 4Nz \\
0 &=& p(\gamma) + 4Ny \\
0 &=& \frac{(2\mu-2\rho-1)p(\rho)}{4N(\mu-\rho)} - \frac{4(\alpha-1)u}
{(2\alpha-1)}
\end{eqnarray}
where $z$ $=$ $A^2B$, $y$ $=$ $A^2C$, $u$ $=$ $A^2D$ with $r(\alpha)$ $=$
$\alpha p(\alpha)/(\mu-\alpha)$, $p(\alpha)$ $=$ $a(\alpha-\mu)/[\pi^{2\mu}
a(\alpha)]$ and $a(\alpha)$ $=$ $\Gamma(\mu-\alpha)/\Gamma(\alpha)$.
Therefore eliminating $z$, $y$ and $u$ from (4)-(7) and substituting the
leading order values for the exponents, one deduces
\begin{equation}
\eta_1 ~=~ - \, \frac{\Gamma(2\mu-1)}{\Gamma(1-\mu)\Gamma(\mu-1)\mu
\Gamma^2(\mu)} \left[ \frac{(2\mu-1)[(2\mu-1)(\mu-2)+\mu b]}{4(\mu-1)^2}
+ 1 \right]
\end{equation}
where $\eta$ $=$ $\sum_{i=1}^\infty \eta_i/N^i$ and (8) represents an
analytic expresion which can be expanded in $d$ $=$ $2$ $+$ $\epsilon$ or
$d$ $=$ $4$ $-$ $2\epsilon$ to compare with other models. Whilst $\eta_1$ is
$b$-dependent it always must be computed first in any calculation since the
other exponents are expressed as a function of it \cite{6}. In three
dimensions we record
\begin{equation}
\eta_1 ~=~ \frac{4(3b-1)}{3\pi^2}
\end{equation}

To determine the anomalous dimensions of the remaining fields one examines
instead of the $2$-point function, the scaling behaviour of the respective
$3$-point vertices in the critical region based on the method of \cite{17}
adapted to the $4$-fermi model with discrete chiral symmetry in \cite{5}. We
find, however, that the extension of the Gross Neveu model with a continuous
global chiral symmetry reveals a significantly different structure for the
exponents compared to the coupling to the case of a discrete symmetry,
\cite{18}. For example, in arbitrary dimensions
\begin{eqnarray}
\chi_{\sigma \, 1} &=& \chi_{\pi \, 1} ~=~ - ~ \frac{(2\mu-1)(2\mu-1+b)\mu
\Gamma(2\mu-1)}{4\Gamma(\mu+1)\Gamma^2(\mu)\Gamma(2-\mu)} \\
\chi_{A \, 1} &=& - \, \eta_1
\end{eqnarray}
The absence of any $O(1/N)$ correction to the gauge field exponent is merely
a consequence of the $U(1)$ Ward identity of the model which was observed in
\cite{19}. In the pure chiral Gross Neveu model, \cite{8}, $\chi_{\sigma \, 1}$
and $\chi_{\pi \, 1}$ are both zero in contrast to (10). The inclusion of the
$A_\mu$ field has given rise to quantum fluctuations which generate an
anomalous piece at $O(1/N)$. Moreover, in $\beta$ and $\gamma$ the gauge
parameter $b$ cancels in all dimensions so that we have deduced the first of
the gauge independent exponents of this letter.

The advantage of solving (1) via the self consistency approach of [4-7] is that
it is relatively straightforward to push the analysis to the next order,
$O(1/N^2)$. This has been carried out separately for the $O(N)$ case, \cite{4},
and for QED, \cite{20}, where the fermion anomalous dimension has been computed
at $O(1/N^2)$ as well as in the amalgam model where the mixed graphs, which
occur at next order, were computed \cite{21}. We have been able to extend that
analysis to (1). This involves including the higher order graphs in the
Dyson equations and we have illustrated a representative set for each Dyson
equation in fig. 2. There are three higher order two loop graphs for each of
the $\sigma$, $\pi$ and $A_\mu$ equations, whilst there are nine for $\psi$.
Due to the high degree of symmetry present the values of only two of these are
independent. Restricting to the Landau gauge we determine,
\begin{eqnarray}
\eta_2 &=& \tilde{\eta}_1^2 \left[ \frac{(2\mu^2-1)(\mu-1)^2}{2} \hat{\Psi}
+ \frac{(\mu-1)^2(2\mu^2-1)}{4} \left( \frac{5}{(\mu-1)} + \frac{2}{(\mu-2)}
\right) \right. \nonumber \\
&+& \!\!\!\! \left. \frac{(2\mu-1)}{32\mu(\mu-1)} [ (2\mu-1) - (\mu-1)^2
+ 2(\mu-1)^4 (2\mu+1)(2\mu-3)] \right. \nonumber \\
&+& \!\!\!\! \left. \frac{[(2\mu-1)(\mu-2)^2 + 4(\mu-1)^2]}{32\mu} \left(
6\mu^2(\mu-1)(2\mu-1)\left( \hat{\Theta} - \frac{1}{(\mu-1)^2} \right)
\right. \right. \nonumber \\
&-& \!\!\!\! \left. \left. \frac{2\mu(\mu-2)(2\mu-1)}{(\mu-1)}
+ (2\mu-1)^2(\mu-2) + 4(\mu-1)^2 \right) \right]
\end{eqnarray}
where $\tilde{\eta}_1$ $=$ $\Gamma(2\mu-1)/[\Gamma(2-\mu)\Gamma^2(\mu)
\Gamma(\mu+1)]$, $\hat{\Theta}(\mu)$ $=$ $\psi^\prime(\mu-1)$ $-$
$\psi^\prime(1)$ and $\hat{\Psi}(\mu)$ $=$ $\psi(2\mu-2)$ $+$ $\psi(3-\mu)$ $-$
$\psi(1)$ $-$ $\psi(\mu)$ and $\psi(x)$ is the logarithmic derivative of the
$\Gamma$-function. The result (12) represents the first $O(1/N^2)$ expression
available for (1) and in three dimensions,
\begin{equation}
\eta_2 ~=~ \frac{4[27\pi^2+14]}{27\pi^4}
\end{equation}

To complete the $O(1/N)$ analysis of (1) we need to compute the corrections to
the exponent $\lambda$ $=$ $\mu$ $-$ $1$ $+$ $O(1/N)$, which has been studied
in other models, \cite{4,16,18}. The first step in the method to achieve this
involves considering corrections to the asymptotic scaling functions (2) of the
form
\begin{eqnarray}
\psi(x) &\sim& \frac{A{x \!\!\! /}}{(x^2)^\alpha}[ 1 + A^\prime (x^2)^\lambda]
{}~~,~~ \sigma(x) ~\sim~ \frac{B}{(x^2)^\beta} [1+B^\prime (x^2)^\lambda]
\nonumber \\
\pi(x) &\sim& \frac{C}{(x^2)^\gamma} [1+C^\prime(x^2)^\lambda] \\
A_{\mu\nu}(x) &\sim& \frac{D}{(x^2)^\rho} \left[ \eta_{\mu\nu}
+ \frac{2\rho(1-b)}{(2\mu-2\rho-1+b)} \frac{x_\mu x_\nu}{x^2} \right.
\nonumber \\
&+& \left. D^\prime (x^2)^\lambda \left( \eta_{\mu\nu}
+ \frac{2(\rho-\lambda)(1-b)}{(2\mu-2\rho+2\lambda-1+b)}\frac{x_\mu x_\nu}{x^2}
\right) \right] \nonumber
\end{eqnarray}
within the critical Dyson equations, where $A^\prime$, $B^\prime$, $C^\prime$
and $D^\prime$ are new amplitudes. This procedure is more laborious than for
the original formulation of \cite{7} since, as was noted in \cite{16}, one has
to include corrections coming from the $\sigma$, $\pi$ and $A_\mu$ two loop
self energy graphs, where there is an $(x^2)^\lambda$ insertion on the
respective lines. Omitting these pieces would result in an erroneous analysis.
To determine the $O(1/N)$ correction to $\lambda$ the consistency equation
based on (14) decouple into a set for $\eta_2$ already considered and a set
involving $A^\prime$, $B^\prime$, $C^\prime$ and $D^\prime$. To ensure this is
a consistent set of equations, the determinant of the matrix formed by these
amplitudes as basis vectors must vanish. With the explicit values of the
$2$-loop integrals calculated in \cite{18}, the determinant yields two
solutions. We discard the trivial one $\lambda_1$ $=$ $0$, in favour of
\begin{equation}
\lambda_1 ~=~ - \, \half (2\mu-1)(2\mu^2-2\mu+1)\tilde{\eta}_1
\end{equation}
and we have checked that the $b$-depedendence has cancelled.

We finish the analysis by restricting our arbitrary dimensional work to three
dimensions and include the term, \cite{11,12},
\begin{equation}
\theta \epsilon_{\mu\nu\sigma}A^\mu\partial^\nu A^\sigma
\end{equation}
in (1), where we work in euclidean space, $\epsilon_{123}$ $=$ $1$ and
$\theta$ is the Chern Simons coupling constant. (The $(F_{\mu\nu})^2$ term of
(1) is now omitted and the transverse contribution to the $A_\mu$ propagator
will be generated in the large $N$ expansion through the relevant graphs of
fig. 1.) The aim will now be to deduce the $\theta$-dependence of the exponents
calculated at $O(1/N)$ due to the potential relation of four fermi models to
high $T_c$ superconductivity. As was noted in \cite{18} one advantage of
calculating with (16) in the large $N$ self consistency approach of \cite{6,7}
is that the spacetime dimension is fixed and therefore one does not need to
extend the definition of $\epsilon_{\mu\nu\sigma}$ to $d$-dimensions as one
would in a perturbative analysis using dimensional regularization, \cite{22}.
The only consequence that (16) has on the scaling forms is the appearance of an
overall factor $(1+\theta^2)^{-1}$ in the scaling form of the gauge field two
point function $A_{\mu\nu}^{-1}(x)$, \cite{18}. With this observation it is
straightforward to repeat the $O(1/N)$ analysis for $b$ $\neq$ $0$ in $d$ $=$
$3$ to find
\begin{eqnarray}
\eta_1 ~=~ \frac{4[3b-1+\theta^2]}{3\pi^2(1+\theta^2)} &,&
\chi_{\sigma \, 1} ~=~\chi_{\pi \, 1} ~=~ - \, \frac{4(b+2)}{\pi^2(1+\theta^2)}
\nonumber \\
\chi_{A \, 1} ~=~ - \, \eta_1 &,&
\lambda_1 ~=~ - \, \frac{8(\theta^2+5)}{3\pi^2(1+\theta^2)}
\end{eqnarray}
where each exponent is a monotonic function of $\theta$. Similar $\theta$
dependent exponents have been obtained at leading order in $1/N$ in other
models, \cite{19,23,24}. As a check on the correctness of (17), they agree with
the earlier $\theta$ $=$ $0$ values, (9), (10) and (15), and as $\theta$
$\rightarrow$ $\infty$ one recovers the previous results of \cite{8} since then
the gauge field decouples from (1) and (16) where clearly one is left simply
with the pure Nambu--Jona-Lasinio model. This latter limit has also been
examined in other models, such as the bosonic $\sigma$ model on $CP(N)$
\cite{23,24}, and the same decoupling feature of the gauge field is observed
there as well.

We conclude with several remarks. First, we have solved the gauged
Nambu--Jona-Lasinio model or chiral Gross Neveu model at leading order large
$N$ by deducing the gauge independent exponents of the model. As these
expressions are in arbitrary dimensions they will be of use in determining
agreement with the $\epsilon$-expansion of the explicit perturbative
renormalization group functions of the theory. By including a Chern Simons
term to reveal non-trivial $\theta$ dependence, these results will be useful
in the experimental situation of differentiating between the nature of the
four fermi interaction since the $\theta$ dependence of our results, (17),
differ substantially from those of \cite{18}. Finally, it is worth pointing out
that the conformal methods we have used here to deduce exponents are much more
efficient in terms of computing integrals than the conventional large $N$
methods and, moreover, have the advantage of allowing one to easily probe the
quantum structure to next to leading order.

\newpage
\noindent
{\Large {\bf Figure Captions.}}
\begin{description}
\item[Fig. 1.] Leading order skeleton Dyson equations.
\item[Fig. 2.] Basic structure of graphs for $O(1/N^2)$ corrections.
\end{description}

\newpage
\begin{thebibliography}{99}
\bibitem{1} W.A. Bardeen, C.T. Hill \& M. Lindner, Phys. Rev. {\bf D41} (1990),
1647.
\bibitem{2} A. Hasenfratz, P. Hasenfratz, K. Jansen, J. Kuti \& Y. Shen, Nucl.
Phys. {\bf B365} (1991), 79.
\bibitem{3} D. Gross \& A. Neveu, Phys. Rev. {\bf D10} (1974), 3235; Y. Nambu
\& G. Jona-Lasinio, Phys. Rev. {\bf 122} (1961), 345.
\bibitem{4} J.A. Gracey, Int. J. Mod. Phys. {\bf A7} (1991), 395; 2755(E).
\bibitem{5} J.A. Gracey, Phys. Lett. {\bf 297B} (1992), 293.
\bibitem{6} A.N. Vasil'ev, Yu.M. Pis'mak \& J.R. Honkonen, Theor. Math. Phys.
{\bf 46} (1981), 157.
\bibitem{7} A.N. Vasil'ev, Yu.M. Pis'mak \& J.R. Honkonen, Theor. Math. Phys.
{\bf 47} (1981), 291.
\bibitem{8} J.A. Gracey, `The Nambu--Jona-Lasinio model at $O(1/N^2)$',
LTH-308 to appear in Phys. Lett. B.
\bibitem{9} A. Kovner \& D. Eliezer, Int. J. Mod. Phys. {\bf A7} (1992), 2755.
\bibitem{10} R. Jackiw \& S. Templeton, Phys. Rev. {\bf D23} (1981), 2291.
\bibitem{11} J. Schonfeld, Nucl. Phys. {\bf B185} (1981), 157.
\bibitem{12} S. Deser, R. Jackiw \& S. Templeton, Phys. Rev. Lett. {\bf 48}
(1982), 975.
\bibitem{13} E. Witten, Commun. Math. Phys. {\bf 121} (1989), 351.
\bibitem{14} I. Dzyaloshinsky, A.M. Polyakov \& P. Wiegman, Phys. Lett. {\bf
121A} (1987), 112.
\bibitem{A} A. Blasi, N. Maggiore \& S.P. Sorella, Phys. Lett. {\bf 285B}
(1992), 54.
\bibitem{15} S. Hands, A. Koci\'{c} \& J.B. Kogut, Ann. Phys. {\bf 224} (1993),
29.
\bibitem{16} J.A. Gracey, `Algorithm for computing the $\beta$-function of
quantum electrodynamics in the large $N_{\! f}$ expansion', LTH-294 to appear
in Int. J. Mod. Phys. A.
\bibitem{17} A.N. Vasil'ev \& M.Yu. Nalimov, Theor. Math. Phys. {\bf 55}
(1983), 423; {\bf 56} (1983), 643.
\bibitem{18} J.A. Gracey, Ann. Phys. {\bf 224} (1993), 275.
\bibitem{19} J.A. Gracey, J. Phys. {\bf A25} (1992), L109.
\bibitem{20} J.A. Gracey, Mod. Phys. Lett. {\bf A7} (1992), 1945.
\bibitem{21} J.A. Gracey, J. Phys. {\bf A26} (1993), 1431.
\bibitem{22} W. Chen, G.W. Semenoff \& Y.S. Wu, Phys. Rev. {\bf D44} (1991),
R1625; L.V. Avdeev, G.V. Grigorev \& D.I. Kazakov, Nucl. Phys. {\bf B382}
(1992), 561.
\bibitem{23} S.H. Park, Mod. Phys. Lett. {\bf A7} (1992), 1579.
\bibitem{24} G. Ferretti \& S.G. Rajeev, Mod. Phys. Lett. {\bf A7} (1992),
2087.
\end{thebibliography}
\end{document}